\documentclass[journal]{IEEEtran}
\usepackage{cite}
\usepackage{amsmath,amssymb,amsfonts}
\usepackage{graphicx,epstopdf}
\usepackage{textcomp}
\usepackage{xcolor}
\usepackage{url}
\usepackage{multirow}
\usepackage{stfloats}

\begin{document}
\bstctlcite{IEEEexample:BSTcontrol}
\title{Evaluating Large Language Models for Automatic Register Transfer Logic Generation
via High-Level Synthesis}

\author{Sneha~Swaroopa,~\IEEEmembership{Graduate Student Member,~IEEE,}
        Rijoy~Mukherjee,~\IEEEmembership{Graduate Student Member,~IEEE,}
        Anushka~Debnath,
        and~Rajat~Subhra~Chakraborty,~\IEEEmembership{Senior Member,~IEEE}

\thanks{S. Swaroopa, R. Mukherjee and R. S. Chakraborty are with the Dept. of Computer Science and Engineering, Indian Institute of Technology, Kharagpur, West Bengal, India 721302. Email: swaroopasneha202@kgpian.iitkgp.ac.in, rijoy.mukherjee@iitkgp.ac.in, rschakraborty@cse.iitkgp.ac.in.}
\thanks{A. Debnath is with the Dept. of Computer Science and Engineering, National Institute of Technology, Durgapur, West Bengal, India 713209. Email: anushkadebnath77777@gmail.com.}
\thanks{Manuscript received XXXXXXXX; revised XXXXXXX}}

\markboth{Journal of \LaTeX\ Class Files,~Vol.~14, No.~8, August~2015}%
{Shell \MakeLowercase{\textit{et al.}}: Bare Demo of IEEEtran.cls for IEEE Journals}

\maketitle

\begin{abstract}
The ever-growing popularity of large language models (LLMs) has resulted 
in their increasing adoption for hardware design and verification. Prior research 
has attempted to assess the capability of LLMs to automate digital hardware design 
by producing superior-quality Register Transfer Logic (RTL) descriptions, particularly
in Verilog. However, these tests have revealed that Verilog code production using LLMs 
at current state-of-the-art lack sufficient functional correctness to be practically
viable, compared to automatic generation of programs in general-purpose programming 
languages such as C, C++, Python, etc. 
With this as the key insight, \textbf{in this paper we assess the performance of a 
two-stage software pipeline for automated Verilog RTL generation: LLM 
based automatic generation of annotated C++ code suitable for high-level 
synthesis (HLS), followed by HLS to generate Verilog RTL.} 
We have benchmarked the performance of our proposed
scheme using the open-source \emph{VerilogEval} dataset, for four different 
industry-scale LLMs, and the \emph{Vitis HLS} tool. 
Our experimental results demonstrate that our two-step technique substantially
outperforms previous proposed techniques of direct Verilog RTL generation by 
LLMs in terms of average functional correctness rates, reaching 
score of 0.86 in $pass@1$ metric. 
\end{abstract}


\IEEEpeerreviewmaketitle

\section{Introduction}
Digital hardware design using a Hardware Description Language (HDL) like
Verilog has traditionally been a labor-intensive and challenging endeavor.
It warrants significant expertise, and is error-prone, necessitating intense
validation efforts for large industry-scale designs. Historically, digital 
hardware design using HDLs has been a specialized skill-set, and far lesser
number of expert designers using HDLs like Verilog exist than expert programmers
in mainstream programming languages such as C, C++, Java, Python, etc.

Naturally, there is an increasing interest in developing more accessible 
approaches for producing HDL, intending to simplify design and testing processes. 
This would make constructing functionally correct hardware easier for those 
with pre-existing software development experience in mainstream programming
languages. Towards this end, High-level synthesis (HLS) has emerged and
matured over the last few decades, as a paradigm that helps software developers 
to design correct-by-construction hardware, by directly translating code from 
a high-level programming language like C or C++ (two most common options), to a 
target HDL like Verilog~\cite{cong2011hls}. For instance, \emph{Vitis HLS}~\cite{VitisHLS} 
and \emph{Bambu}~\cite{ferrandi2021bambu} are free-of-cost software tools that 
facilitate rapid hardware logic prototyping by generating RTL in HDL, based on 
hardware specifications in (supersets of) C, C++, and System C. 

Recent efforts have been aimed at elevating the abstraction level and ease of
digital hardware design even further, where the goal is to allow designers 
to articulate functional specifications of the logic to be described in a 
natural language (NL)~\cite{Pearce2020DAVE}. State-of-the-art Large 
Language Models (LLMs) have the potential to revolutionize digital hardware
design and validation methodology, by automatically generating correct-by-construction 
circuit descriptions in HDL, and HDL code for hardware validation (``testbench"),
when prompted by the functional description of the circuit in a natural language, 
e.g. English. 

However, recent studies reported by multiple research groups to evaluate the potential of
LLMs at the current state-of-the-art in producing functionally correct HDL descriptions
(mostly RTL in Verilog) have identified limitations regarding their ability 
to generate correct and complete Verilog~\cite{liu2023verilogeval}. Input 
``prompt engineering'' targeting LLMs for automatic Verilog generation also 
lacks structure and is sometimes ambiguous. Another constraint 
is that currently, LLMs cannot perform top-down digital design, necessitating (relatively)
much more laborious bottom-up design~\cite{liu2023verilogeval}.
LLMs often generate HDL descriptions without ensuring functional correctness at
the hardware level and usually do not perform any design space 
exploration~\cite{chang2023chipgpt}. The main reason for
such low effectiveness of LLMs in HDL generation tasks is the relatively small
amount of open-source codebases in HDLs like Verilog and VHDL available for 
training the LLMs. These limitations currently prevent 
seamless integration 
of LLMs into EDA workflows.
\begin{figure*}[!t]
  \centering
  \includegraphics[scale=0.20]{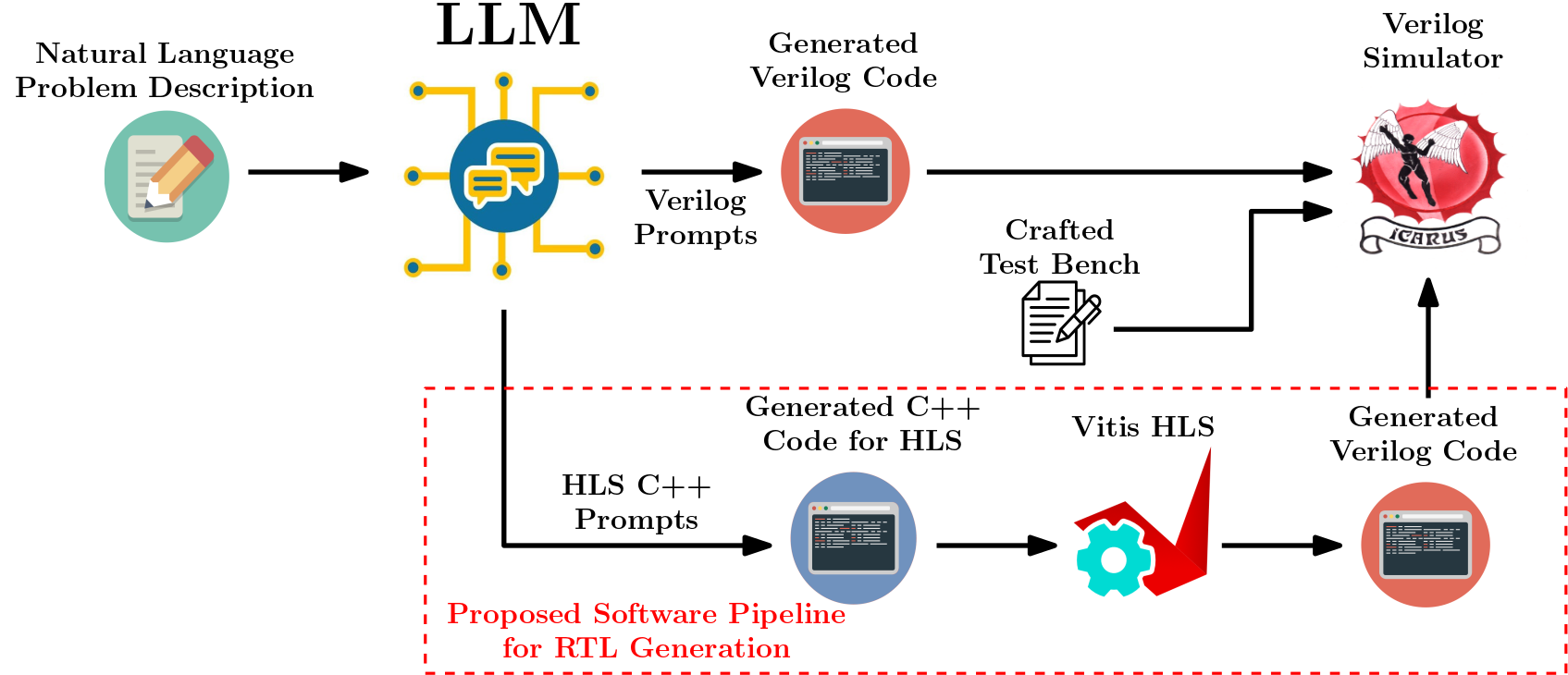}
  \caption{Comparison of existing RTL Generation via LLM 
  with the proposed software pipeline with HLS.}
  \label{fig:workflow_llm_hls}
\end{figure*}

On the other hand, advances in LLMs have led to phenomenal results on 
automatic code generation tasks~\cite{roziere2024codellama}. This is made
possible by the availability of large collections of open-source codebases
in common general-purpose programming languages, which can be used for 
fine-tuned training of the LLMs. 
LLMs have exhibited unprecedented success in diverse tasks like suggesting 
code snippets (``copilots"), solving complex algorithms, and explaining 
programming concepts. \textbf{Therefore, combining HLS which allows 
logic design through higher-level programming interfaces, with the emerging 
LLMs, represents progress toward the vision of automatic and error-free 
hardware logic design via natural language interaction.}

In this paper, we explore the capabilities of LLM in generating sophisticated
C++ programs suitable for HLS processing, starting with natural language problem 
descriptions. We then use a HLS tool to convert this generated C++ code to Verilog
RTL. Since the reliability of hardware design is a key factor, we enable the
evaluation of the generated final Verilog code through a robust validation procedure~\cite{VerilogEvalGit}. Fig.~\ref{fig:workflow_llm_hls} 
compares the proposed software pipeline of LLM-assisted automated hardware logic 
design, with that of the existing works. We experimented with four
different well-known industry-standard LLMs (\texttt{gpt-3.5 turbo}~\cite{gpt-3.5}, \texttt{gpt-4o}~\cite{gpt-4o}, 
\texttt{Claude 3 Haiku}~\cite{haiku}, \texttt{Claude 3.5 Sonnet}~\cite{sonnet}), 
and \emph{Vitis HLS}.
We use a subset of the \emph{VerilogEval} dataset~\cite{liu2023verilogeval},
comprising 70 problems sourced from the \emph{HDLBits} platform, to evaluate 
the approaches (the proposed software pipeline vs. existing techniques). 
We call this the \textbf{HLSEval} dataset, which establishes the superior accuracy 
(up to $pass@1$ value of 0.86) for
our proposed approach, over existing LLM-assisted automatic Verilog RTL generation. 
We have made the \textbf{HLSEval} benchmarks and LLM scripts open-source 
on Github~\cite{HLSEvalGit}.

Next, we describe some related work on automated RTL generation and validation
using LLMs.

\section{LLM-based Automated Digital Hardware Design and Validation}
With the escalating popularity of LLMs, several research works have aimed 
at harnessing their power to generate HDL (primarily Verilog) code, 
to automate digital hardware design and validation.
Both~\cite{dehaerne2023deeplearningframeworkverilog} and~\cite{Thakur2024VeriGen} 
fine-tuned the open-source \texttt{CodeGen} model~\cite{nijkamp2022codegen} with 
the training data consists of Verilog codes from GitHub and machine-readable 
Verilog code snippets from electronic copies of Verilog textbooks. 
Evaluation of curated tasks concluded that the resulting \texttt{VeriGen} 
model~\cite{Thakur2024VeriGen} delivered functionally accurate code only 
$41.9\%$ of the time. 
In~\cite{liu2023verilogeval}, the authors designed a benchmarking framework 
to assess LLM performance in Verilog code generation tasks in hardware design. 
They also introduced a robust, comprehensive dataset called \textit{VerilogEval}, 
comprising 156 problems sourced from HDLBits~\cite{hdlbits}, 
an online Verilog learning platform. The evaluation demonstrated 
the Verilog code generation capabilities of pre-trained language models 
like \texttt{gpt-3.5}~\cite{gpt-3.5} and \texttt{gpt-4}~\cite{gpt-4}, 
and the fine-tuned \texttt{VeriGen} model, which incidentally achieved 
the best performance. However, the improvement wasn't much from the 
numbers achieved previously. To improve the quality of Verilog generation by 
pre-trained models, authors in~\cite{thakur2024autochip} developed a workflow 
with a chain of feedback mechanisms called \textit{AutoChip} to prompt the 
LLMs for iterative hardware development. 
Authors in~\cite{chang2024dataneedfinetuningllms} introduced \texttt{ChipGPT-FT}, 
a Verilog design model, by fine-tuning the \texttt{Llama-2}~\cite{touvron2023llama2} 
model with the Verilog dataset augmented from GitHub. When compared with performance of 
\texttt{VeriGen}~\cite{Thakur2024VeriGen} on similar benchmarks, \texttt{ChipGPT-FT} 
increases functional correctness of the generated code from 
$58.8\%$ to $70.6\%$.

Another branch of work targeting LLMs has focused on the quality of the generated 
Verilog code regarding creativity, optimization, resource utilization and 
flexibility~\cite{chang2023chipgpt, delorenzo2024makecount, delorenzo2024creativevaleval, 2023BlockloveChip-Chat}. 
Other prominent works in this domain have focused on using LLMs for hardware testing~\cite{qiu2024autobench, blocklove2024evaluatingllmshardwaredesign}, 
bug fixing~\cite{Ahmad2024BugCodeFixes}, generating security assertions~\cite{paria2023divas, Kande2024Assertions, meng2023assurance} and formal verification of
RTL~\cite{orenesvera2023formal}. Works such as~\cite{Wu2024ChatEDA} 
and~\cite{liu2024chipnemo} use LLM to enhance hardware design productivity 
by providing a conversational interface like an engineering chatbot that 
helps in task planning, EDA script generation, bug analysis and task execution 
through the invocation of EDA tool APIs.

Very few studies have been conducted in the LLM realm regarding high-level synthesis. 
Authors in~\cite{collini2024c2hlsc} proposed a workflow called \textit{C2HLSC}, 
enabling seamless refactoring of a traditional C program to HLS-compatible 
C, to produce optimized hardware architectures. In~\cite{Fu2023GPT4AIGChip}, 
authors developed a framework called \textit{GPT4AIGChip}, to design an 
AI accelerator, starting from human language specifications. Specifically, it 
used \texttt{gpt-4}~\cite{gpt-4} to design the AI accelerator by prompting 
for decoupled HLS C++ modules. Though the authors investigated the viability 
of HLS with the conclusion that current LLMs struggle with understanding 
long dependencies, the experimentation was not thorough and comprehensive 
enough to account for various types of tasks frequently seen for hardware design 
in HLS. 
\begin{figure}[!t]
  \centering
  \includegraphics[trim=4cm 28.7cm 4cm 9cm, clip, scale=0.30]{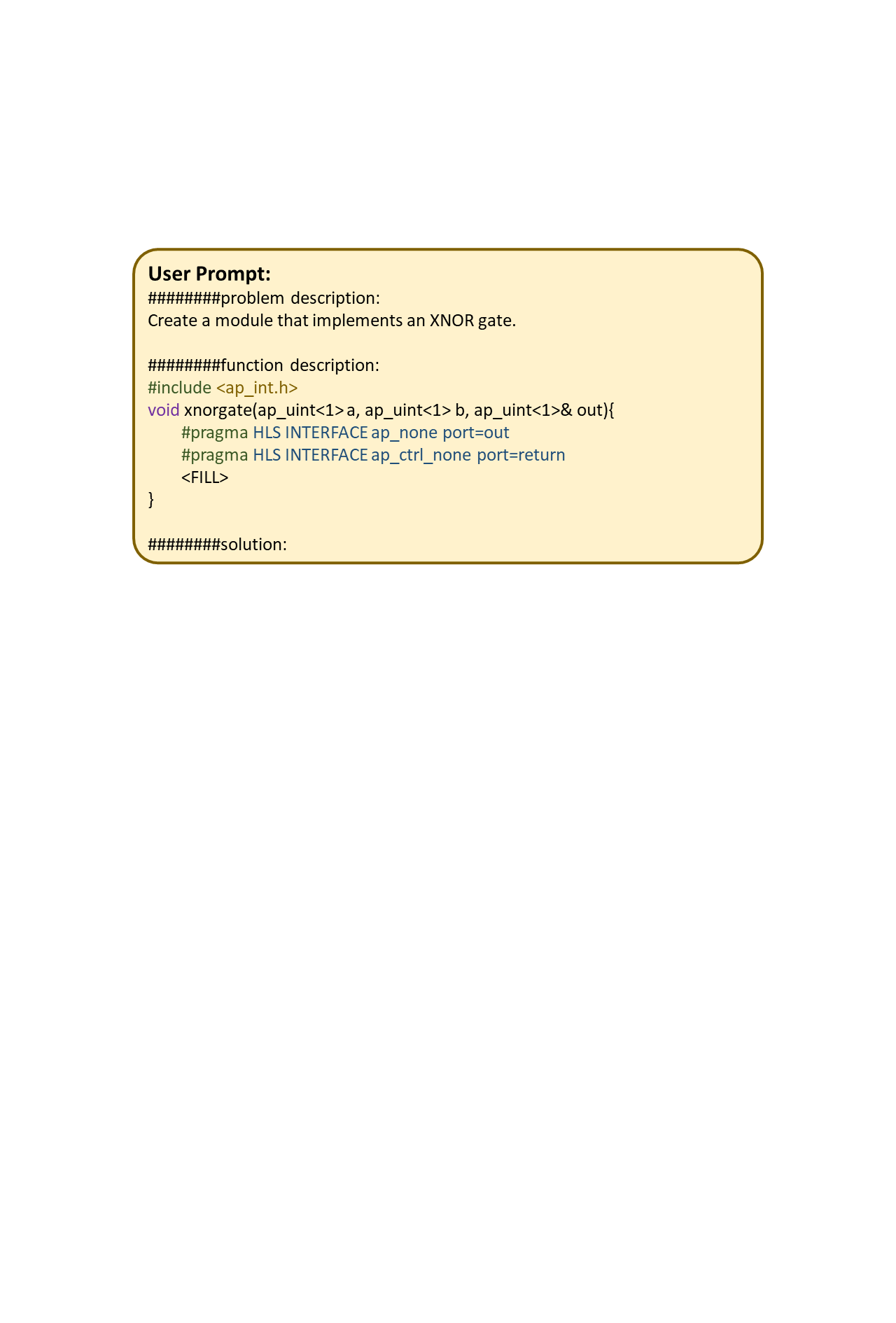}
  \caption{Problem desciption of \texttt{xnor} in \textbf{HLSEval}.}
  \label{fig:xnor_prompt}
\end{figure}
\begin{figure}[!b]
  \centering
  \includegraphics[trim=4cm 19.5cm 4cm 9cm, clip, scale=0.30]{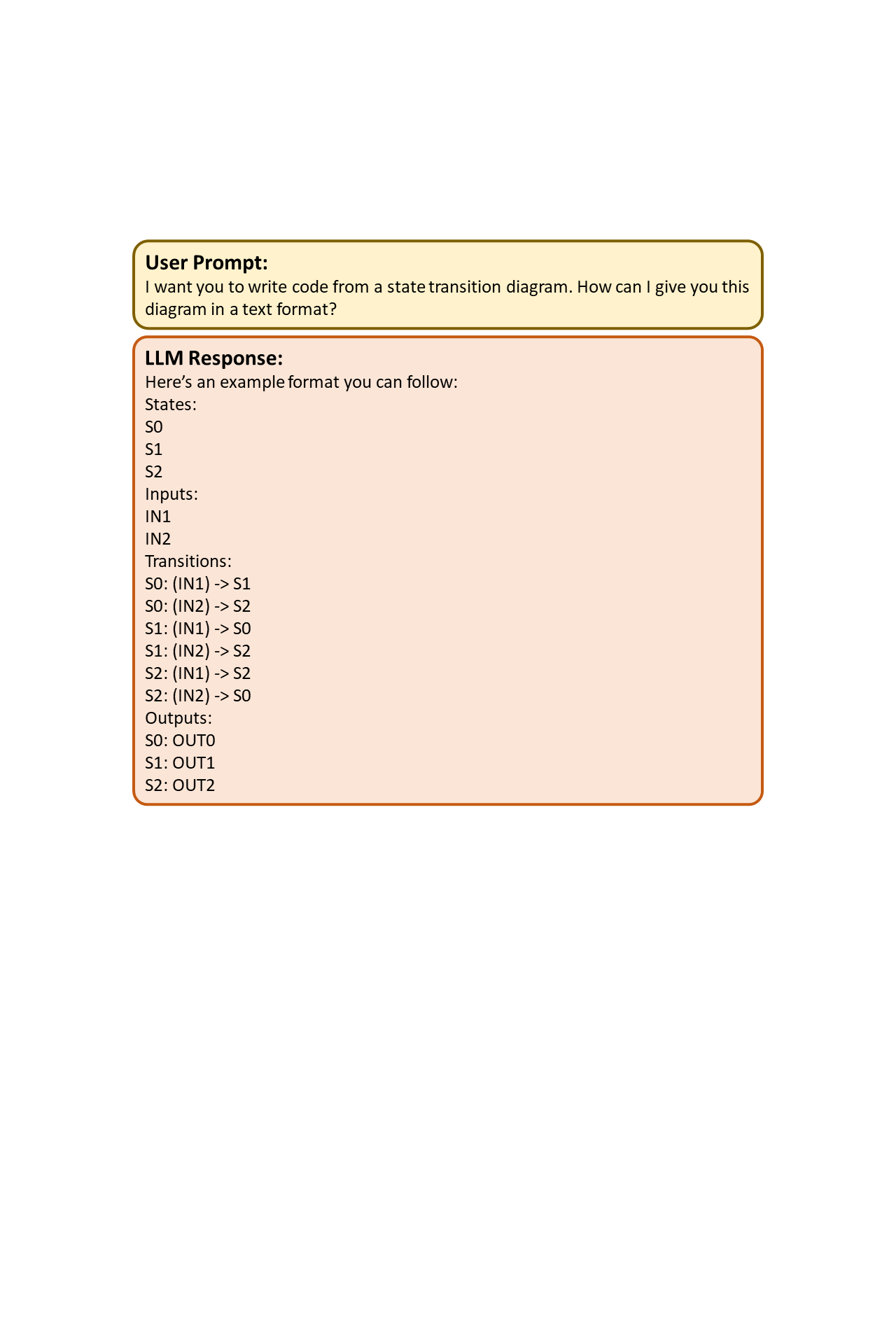}
  \caption{\texttt{gpt-3.5 turbo} response for Finite State Machine (FSM) specifications.}
  \label{fig:state_machine_prompt}
\end{figure}
\begin{figure}[!b]
  \centering
  \includegraphics[trim=4cm 26.8cm 4cm 9cm, clip, scale=0.30]{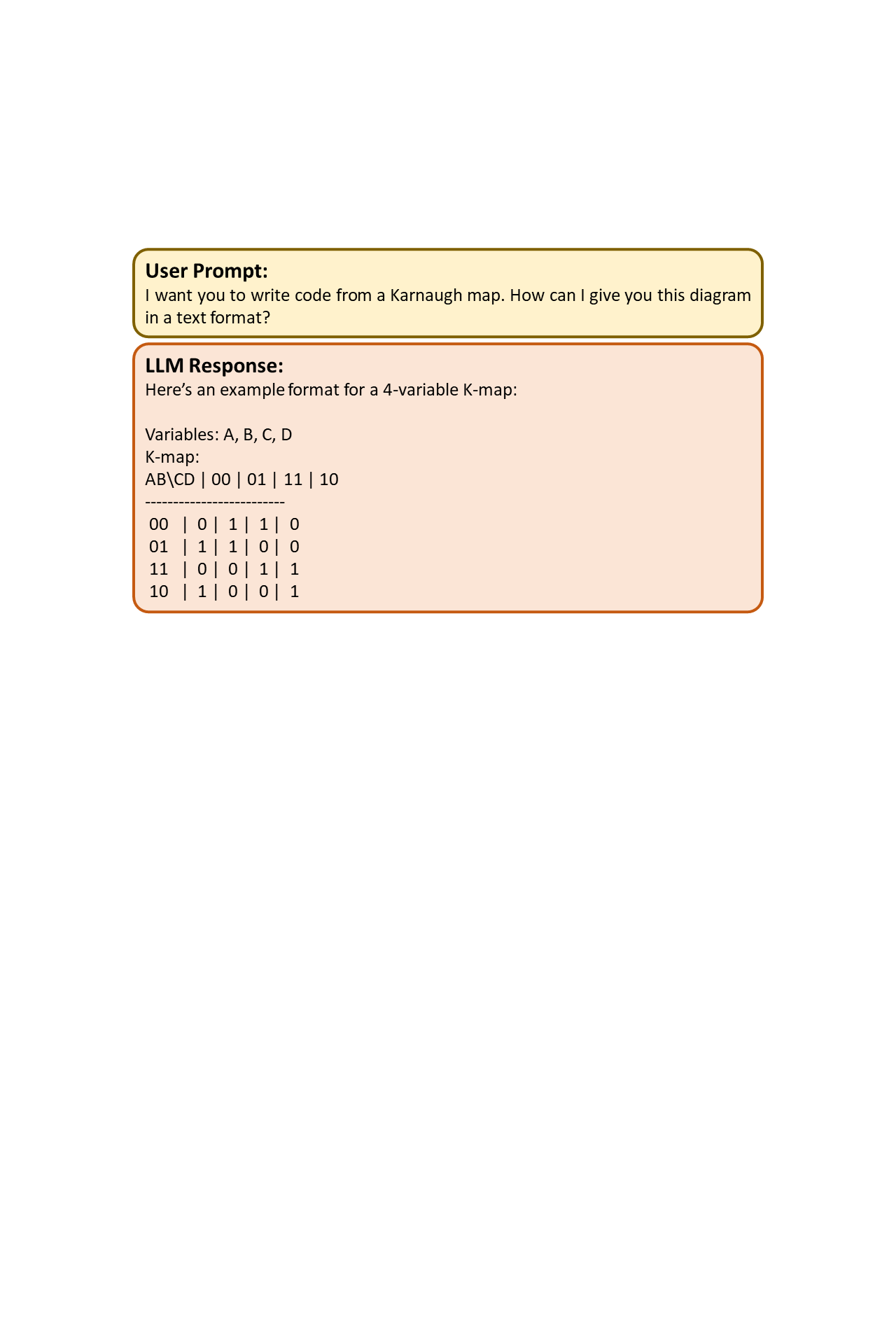}
  \caption{\texttt{gpt-3.5 turbo} response for Karnaugh map (K-map) specification.}
  \label{fig:kmap_prompt}
\end{figure}
\section{Proposed Software Pipeline for Automated Verilog Generation
and Its Validation}
In this section, we discuss in detail the \emph{HLSEval} dataset;
our proposed software pipeline for automated Verilog generation, and its
evaluation technique.

\subsection{The \textbf{HLSEval} Dataset}
We assess the functional correctness of the proposed hardware design pipeline 
on problems selected from the \textbf{VerilogEval}~\cite{liu2023verilogeval} dataset.
\emph{VerilogEval} encompasses diverse coding tasks, ranging from simple combinational 
circuits to complex finite state machines. Our proposed \textbf{HLSEval} dataset 
consists of concise hardware design tasks that demand problem-solving skills in circuit 
optimization, boolean logic reduction, state transitions, and more. 
We only consider a subset of 70 combinational logic circuits 
for this evaluation process. Sequential circuits 
are realized in HLS in the form of \emph{streams}~\cite{VitisHLS}, 
whereas in RTL, they are realized
with one or more user-defined clock signal(s) controlling a module.
Thus, a fair comparison between our proposed two-step approach with 
that of direct LLM-generated Verilog is difficult due to contrasting 
underlying concepts~\cite{Cong2022FPGASuccesses, VitisHLS}.
So, we did not include those 62 sequential problems. 
Remaining problems in \emph{VerilogEval} were related to 
bug-finding in Verilog codes and text-based question answers, 
and are hence unrelated to our scope.

For \textbf{HLSEval}, the problem descriptions from the \textbf{VerilogEval} 
are converted into a schema suitable for HLS C++ prompting. A sample for problem 
statement \texttt{xnor} is shown in Fig.~\ref{fig:xnor_prompt}. It has two 
parts: 1) the problem description representing the hardware generation task 
in simple English, and, 2) the function description providing 
a snapshot of the function signature we are expecting from the LLM. 
Representing Finite State Machines (FSMs) and Karnaugh maps (K-maps) 
into textual forms was a challenge, and for this, we queried 
\texttt{gpt-3.5 turbo}~\cite{gpt-3.5} as shown in 
Fig.~\ref{fig:state_machine_prompt} and Fig.~\ref{fig:kmap_prompt}. We adopted the 
textual structure as returned by \texttt{gpt-3.5 turbo}, for 
further prompting by LLM.
\begin{figure}[!t]
  \centering
  \includegraphics[trim=4cm 2.5cm 4cm 0cm, clip, scale=0.30]{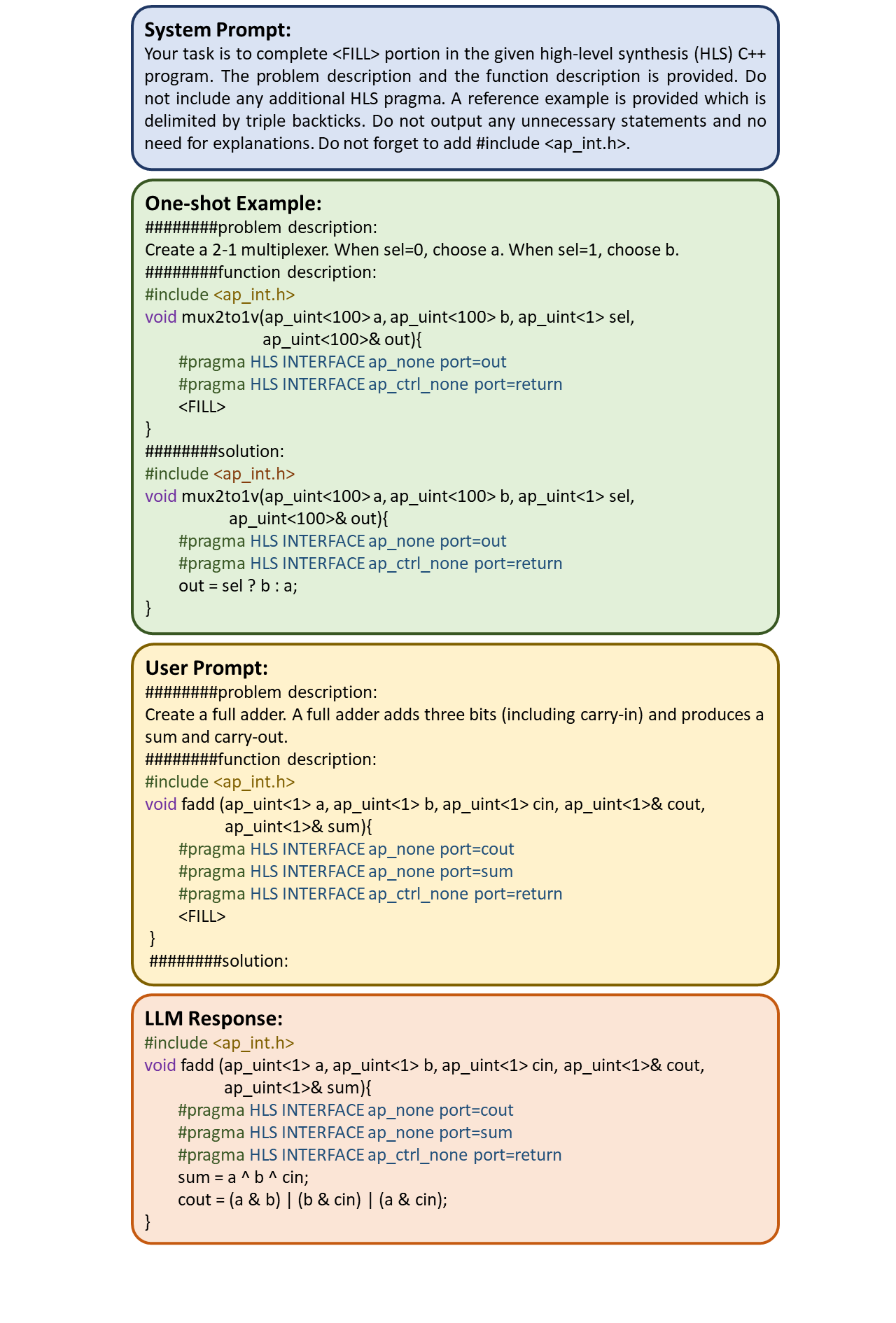}
  \caption{Example of prompting for \texttt{fadd} (full adder) in \textbf{HLSEval}. 
  The logic description includes a description of the problem in natural 
  language, a function description, and a sample one-shot input and output
  definition.}
  \label{fig:prompt_example}
\end{figure}
\begin{figure}[!t]
  \centering
  \includegraphics[trim=4cm 19cm 11.5cm 9cm, clip, scale=0.30]{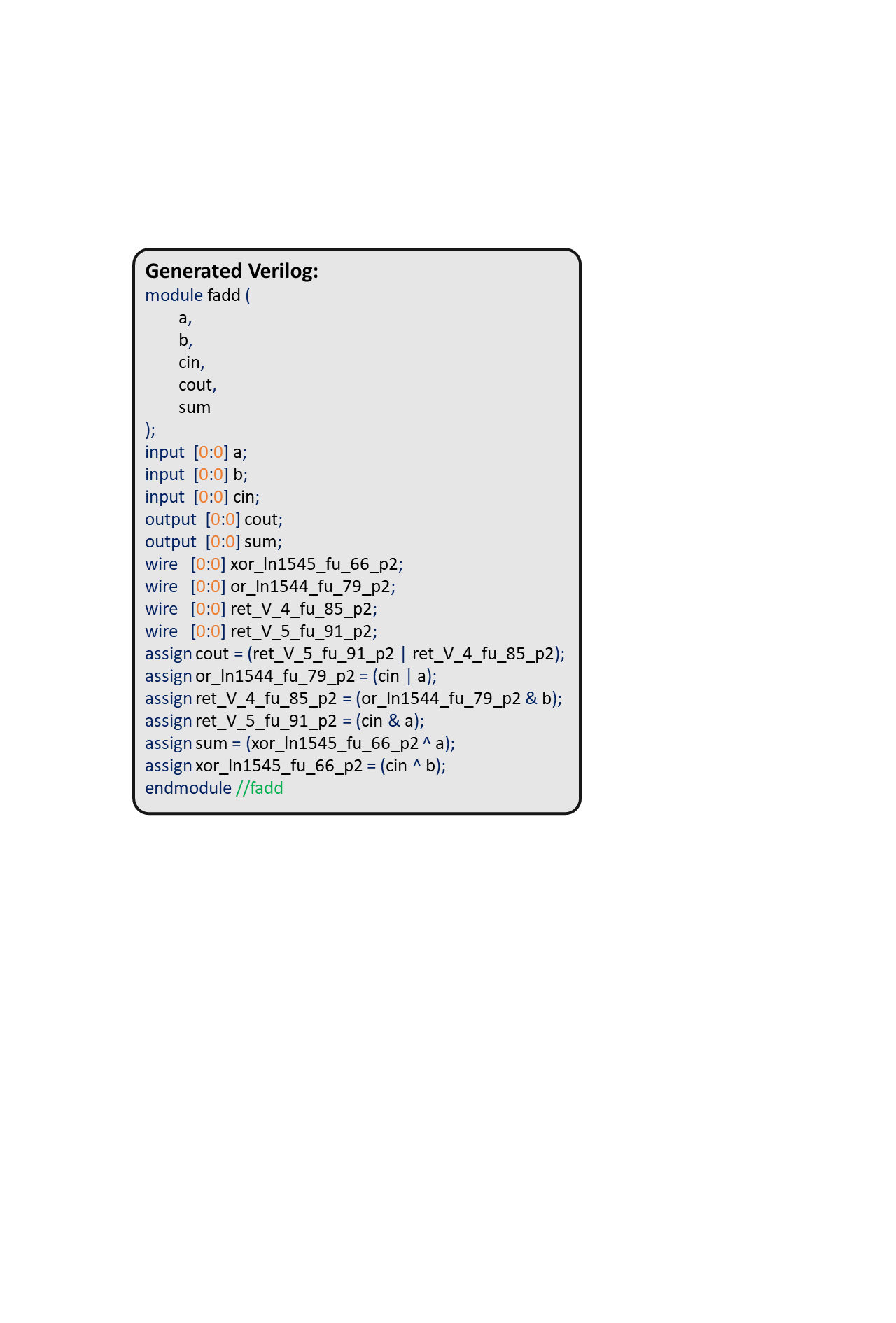}
  \caption{Verilog RTL of \texttt{fadd} generated by \textit{Vitis HLS}.}
  \label{fig:verilog_example}
\end{figure}
\subsection{Proposed Software Pipeline for Automated Verilog Generation}
The first stage of the proposed pipeline, as shown in Fig.~\ref{fig:workflow_llm_hls}, 
consists of the prompting the LLM properly with well-crafted problem 
statements in a natural language (English in our case). 
Fig.~\ref{fig:prompt_example} shows an example of such prompting for the 
problem \texttt{fadd} (a binary full adder). The system prompt is concatenated 
with an ``one-shot" example and user prompt, and sent to the LLM for inference. 
The one-shot example helps to address ambiguity within the problem descriptions, 
particularly when there is a precise requirement for certain attributes in 
the LLM response. The pragmas introduced for each problem are optional. In 
our case, we use it to align the downstream Verilog generated with 
those generated by the existing pipeline of~\cite{liu2023verilogeval} for 
better comparison. Fig.~\ref{fig:prompt_example} also shows the response 
produced by LLM, an annotated HLS-compliant C++ code snippet with the provided function 
signature, and correct functionality. Next, the generated HLS C++ code 
produced by LLM is converted to Verilog using \textit{Vitis HLS}~\cite{VitisHLS}.
Fig.~\ref{fig:verilog_example} shows the generated Verilog for \texttt{fadd} 
by \textit{Vitis HLS}, corresponding to the C++ code in Fig.~\ref{fig:prompt_example}. 
\begin{table*}[!b]
\centering
\caption{Comparison of different models on \textbf{HLSEval} dataset (average $pass@k$ values)}
\label{tab:nl-to-verilog}
\scalebox{1.05}{%
\begin{tabular}{|c|ccc|ccc|}
\hline
\multirow{2}{*}{Model} & \multicolumn{3}{c|}{\textbf{NL to Verilog}}                                  & \multicolumn{3}{c|}{\begin{tabular}[c]{@{}c@{}}\textbf{NL to Verilog via HLS}\\ \textbf{(This Work)}\end{tabular}} \\ \cline{2-7} 
                       & \multicolumn{1}{c|}{$pass@1$} & \multicolumn{1}{c|}{$pass@5$} & $pass@10$ & \multicolumn{1}{c|}{$pass@1$}           & \multicolumn{1}{c|}{$pass@5$}           & $pass@10$          \\ \hline
\texttt{gpt-3.5 turbo}          & \multicolumn{1}{c|}{0.36}   & \multicolumn{1}{c|}{0.53}   & 0.56    & \multicolumn{1}{c|}{0.60}             & \multicolumn{1}{c|}{0.66}             & 0.68             \\ \hline
\texttt{gpt-4o}          & \multicolumn{1}{c|}{0.67}   & \multicolumn{1}{c|}{0.75}   & 0.77    & \multicolumn{1}{c|}{0.80}             & \multicolumn{1}{c|}{0.83}             & 0.84             \\ \hline
\texttt{Claude 3 Haiku}          & \multicolumn{1}{c|}{0.50}   & \multicolumn{1}{c|}{0.63}   & 0.66    & \multicolumn{1}{c|}{0.64}             & \multicolumn{1}{c|}{0.72}             & 0.74             \\ \hline
\texttt{Claude 3.5 Sonnet}          & \multicolumn{1}{c|}{0.70}   & \multicolumn{1}{c|}{0.80}   & 0.83    & \multicolumn{1}{c|}{\textbf{0.86}}             & \multicolumn{1}{c|}{\textbf{0.87}}             & \textbf{0.87}             \\ \hline
\end{tabular}}
\end{table*}
\subsection{Evaluation Methodology for Proposed Technique}
We use the open-source evaluation harness described in~\cite{VerilogEvalGit} to validate 
the Verilog files generated by our pipeline. It compares simulation results between 
the generated Verilog, and ``golden reference" Verilog implementations. It uses the 
open-source \emph{Icarus Verilog}~\cite{icarus} simulator adapted in a sandbox environment 
to safely run the generated Verilog files. We use the popular $pass@k$ 
metric~\cite{chen2021evaluatinglargelanguagemodels} to measure functional correctness
of the generated Verilog RTL, where a problem is considered solved if 
any one of $k$ randomly selected samples pass the unit tests:
\begin{equation}
pass@k := \mathbb{E}_{\text{Problems}} \left[ 1 - \frac{\binom{n-c}{k}}{\binom{n}{k}} \right],
\label{eq:passk}
\end{equation}
where we generate $n \geq k$ samples per problem among which $c \leq n$ 
samples pass testing. In practice, the number of samples $n$ must be sufficiently 
large (at least twice the value of $k$) to produce low-variance estimates for $pass@k$.

Next, we describe comparative experimental evaluation results.
\section{Experimental Results}
For testing, we use four of the best state-of-the-art commercial LLMs, namely, 
\texttt{gpt-3.5 turbo}~\cite{gpt-3.5} and \texttt{gpt-4o}~\cite{gpt-4o} from
\textit{OpenAI}, and \texttt{Claude 3 Haiku}~\cite{haiku} and 
\texttt{Claude 3.5 Sonnet}~\cite{sonnet} by \textit{Anthropic} (the last two
were hosted on Amazon Web Services). We set the parameters to be:
\emph{top p} = $0.95$, temperature = $0.8$, and context length = $2048$ 
for all the experiments. For measuring $pass@k = \{1, 5, 10\}$ we use
Eqn.~\eqref{eq:passk}, with $n = 20$ code completions 
for each problem.
\subsection{English to Verilog RTL: Direct Conversion vs. Two-stage Proposed 
Software Pipeline}
Table~\ref{tab:nl-to-verilog} illustrates the $pass@$ rates for the 
\textbf{HLSEval} tasks using the various LLMs, and compares both 
techniques. \textbf{NL to Verilog} refers to existing 
technique~\cite{liu2023verilogeval}, where LLM generates Verilog 
based on natural language, whereas \textbf{NL to Verilog via HLS} 
refers to our proposed approach where LLM generates HLS-compatible 
C++ based on natural language, that is then converted to Verilog. 
From the study in~\cite{liu2023verilogeval}, more capable and 
larger LLMs generally result in better Verilog code generation 
capability. This observation is consistent with our results 
for the four LLMs, both for the \textbf{NL to Verilog} scheme,
as well as the \textbf{NL to Verilog via HLS} scheme. 
However, \textbf{we obtained significant improvement in $pass@$ 
values} for the proposed \textbf{NL to Verilog via HLS} software
pipeline, compared to the \textbf{NL to Verilog} technique.

\subsection{Current Limitations}
\begin{figure}[!t]
  \centering
  \includegraphics[trim=7.5cm 8.5cm 8.5cm 1cm, clip, scale=0.4]{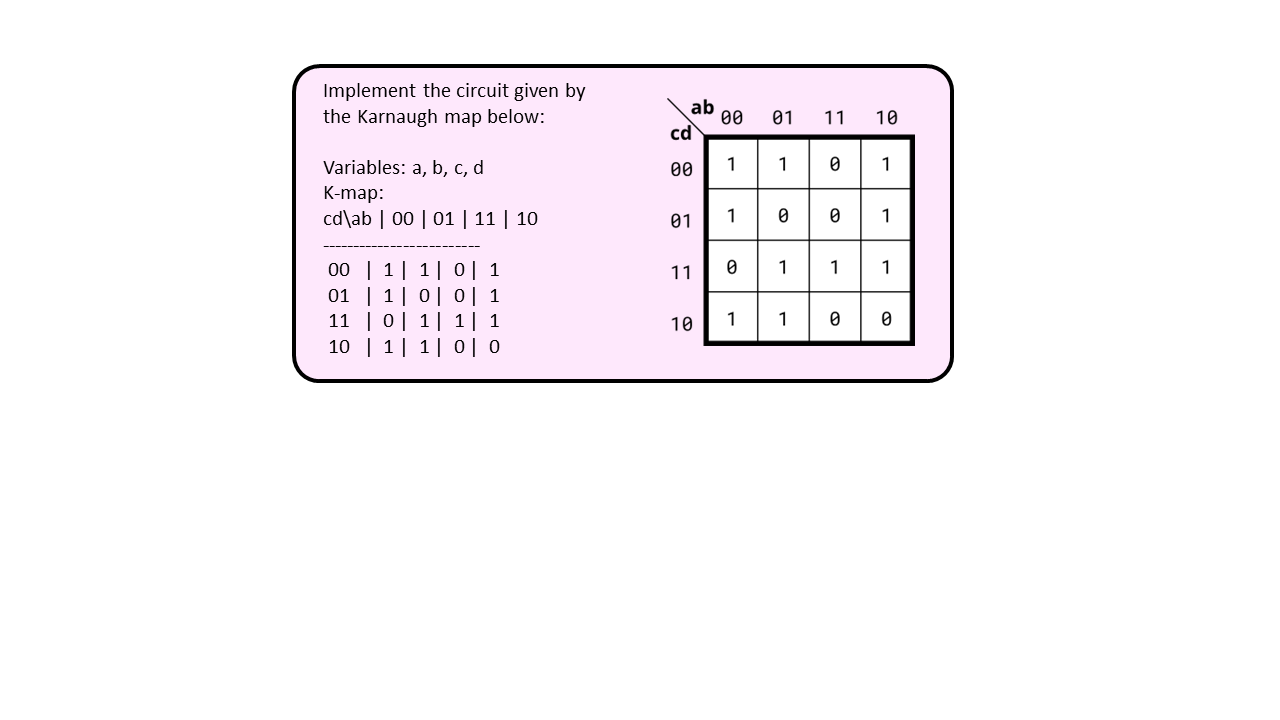}
  \caption{An example of K-map problem.}
  \label{fig:kmap_task}
\end{figure}

One major challenge we found in automated logic design 
from natural language is realising the circuits corresponding to
K-map problems like those of Fig.~\ref{fig:kmap_task}. Even 
though we prompt the LLM with the schema provided in 
Fig.~\ref{fig:kmap_prompt}, none of the four LLMs generated 
correct code. By analyzing the generated (incorrect) Verilog
RTL codes, we found that the LLMs could not minimize 
the K-maps, or correctly identify the minterms or maxterms. 
This could be a future work domain for LLMs to solve, by 
fine-tuning with more data or advanced prompting techniques.

\section{Conclusion}
Researchers have applied LLMs to generate RTL descriptions for
digital hardware from hardware specifications written in natural 
language. This paper proposes an alternative as a software pipeline 
for automatic hardware logic design, where LLMs are prompted to 
produce HLS-compatible C++ code, followed by HLS of the generated
C++ code to Verilog RTL. Detailed experiments on a standard 
benchmark suite demonstrated that Verilog code generation through 
the proposed software pipeline achieves much higher functional
correctness rate.

\bibliographystyle{IEEEtran}
\bibliography{bibfile}

\end{document}